\documentclass[twocolumn,showpacs,preprintnumbers,amsmath,amssymb]{revtex4}
\date{\today}

\usepackage{graphicx}

\newcommand{\rr}{\mbox{\boldmath $r$}}
\newcommand{\EE}{\mbox{\boldmath $E$}}

\newcommand{\hh}{\mbox{\boldmath $h$}}

\newcommand{\OOmega}{\mbox{\boldmath $\Omega$}}
\begin{document}


\title{
Sagnac Effect in Resonant Microcavities
}

\author{Satoshi Sunada
and Takahisa Harayama}
\affiliation{
$^1$Department of Nonlinear Science ATR Wave Engineering Laboratories
2-2-2 Hikaridai Seika-cho Soraku-gun Kyoto 619-0228 Japan\\  
}

\begin{abstract}
%
%
The Sagnac effect in two dimensional (2D) resonant microcavities
is studied theoretically and numerically.
The frequency shift due to the Sagnac effect occurs
as a threshold phenomenon for the angular velocity in a rotating microcavity.
Above the threshold, the eigenfunctions of a rotating microcavity 
become rotating waves while they are standing waves below the threshold. 
\end{abstract}


\maketitle
The Sagnac effect is the phase difference between 
two counter-propagating laser beams in the same ring resonator 
due to rotation, originally introduced by Sagnac in 1913\cite{Post}. 
It has become the basis for the operation of the optical 
gyroscopes such as ring laser gyroscopes and fiber optic gyroscopes 
after the invention of lasers and optical fibers in 1970's
\cite{Crow,Aronowitz,Vali,Ezekiel}
because the phase and frequency difference between clockwise (CW) 
and counter-clockwise (CCW) propagating beams are proportional to the applied 
angular velocity. 
These optical gyroscopes are normally used in airplanes, rockets, 
and ships etc. since they are the most precise rotation velocity sensors   
among any other types of gyroscopes. 

The Sagnac effect had been theoretically derived for the slender waveguides 
like optical fibers or the ring cavities composed of more than three mirrors 
by assuming that the light propagates one-dimensionally 
and the wavelength of the light is much shorter 
than the sizes of the waveguides or the ring cavities\cite{Post,Crow,Lamb2}. 
However, the sizes of the resonant cavities can be reduced to the order 
of the wavelength by modern semiconductor technologies 
\cite{Yamamoto,Optical processes,Nockel Stone}. 
The conventional description of the Sagnac effect is not applicable 
to such small resonant microcavities. Especially, the resonance wave 
functions are standing waves which can never be represented 
by the superposition of counter-propagating waves.
The assumption of the existence of CW and CCW 
waves plays the most important role for the conventional theory 
of the Sagnac effect. 

In this Letter, by perturbation theory typically used in quantum mechanics,  
we show that the Sagnac effect can also be observed 
even in resonant microcavities if the angular velocity of the cavity 
is larger than a certain threshold  
where the standing wave resonance function changes into the rotating wave. 
It is also shown that numerical results of the quadrupole cavity correspond 
very well to the theoretical prediction.  
Theoretical and numerical approaches shown in this Letter 
do not assume that the CW and CCW waves exist 
in the cavity, but the pair of the counter-propagating waves is automatically 
produced by mixing the nearly degenerate resonance wave functions 
due to rotation of the cavity. 

According to the general theory of relativity,  
the electromagnetic fields in a rotating resonant microcavity 
are subject to the Maxwell equations generalized to a non-inertial 
frame of reference in uniform rotation with angular velocity vector 
$\OOmega$\cite{Post,Crow,Lamb2,Landau}. 
By neglecting $O(h^2)$, we obtain the following 
wave equation for the electric field $\EE$,
\begin{eqnarray}
& &
\left(
\dfrac{n^2}{c^2}\dfrac{\partial^2}{\partial t^2}-\nabla^2
\right)\EE
+ 
\nabla(\nabla\cdot \EE) \nonumber \\  
& &+    
\dfrac{1}{c}\dfrac{\partial}{\partial t}
\left[
\nabla\times(\hh \times\EE)   
+ \hh\times(\nabla \times\EE) 
\right]
= 0, 
\end{eqnarray}
where 
$
\hh = \frac{1}{c}
\left(
\rr\times\OOmega
\right).
$
In the above, 
$c$ and $n$ are respectively the velocity of light and the refractive index
inside the cavity.
%
%

In conventional theoretical approach for the Sagnac effect,
the frequency shift of the resonance proportional to the angular 
velocity of the rotating ring cavity is derived from
 assuming that the electric field in Eq.~(1) propagates 
one-dimensionally along the slender optical waveguides\cite{Crow,Lamb2}.  
This method is not applicable to the resonant microcavities 
because the wavelength of the resonance is not much shorter  
than the size of the cavity and the electric field does not 
propagate one-dimensionally. 

Instead, in the case of the 2D resonant microcavity 
perpendicular to angular velocity vector $\OOmega$, 
the resonances can be obtained by solving the following 
stationary wave equation derived from Eq.~(1) 
for the stable oscillation solution, 
\begin{eqnarray}
\left(
{\nabla_{xy}}^2 + n^2 k^2
\right)\psi
-
2ik
\left(\hh\cdot\nabla
\right)\psi
= 0, 
\label{fundeq}
\end{eqnarray}
where the 2D resonant cavity is rotating on $xy$-plane clockwisely,i.e.,  
$\OOmega=(0,0,\Omega)$ and $\Omega>0$. 
We assumed that TM mode of the electric field oscillates as 
$\EE(\rr,t) = (0,0,\psi(\rr)e^{-ick t}+c.c.)$.

For simplicity, we impose the Dirichlet boundary condition 
on the electric field of the resonant microcavity 
in the remainder of this Letter. 

In the case of a microdisk cavity, Eq.~(2) can be solved exactly 
as follows. Eq.~(2) is rewritten in the following form 
in the cylindrical coordinates, 
\begin{eqnarray}
\left[
\dfrac{\partial^2}{\partial r^2}
+
\dfrac{1}{\partial r}\dfrac{\partial}{\partial r}
+
\dfrac{1}{r^2}\dfrac{\partial^2}{\partial \theta^2}
+
2ik\dfrac{\Omega}{c}\dfrac{\partial}{\partial \theta}
+
n^2k^2
\right]\psi
=0. 
\label{r-fundeq}
\end{eqnarray}
One can assume the solution $\psi(r,\theta)$ is given as 
$\psi(r,\theta) = f(r)e^{im\theta}$ where $m$ is an integer, 
and then obtains 
\begin{eqnarray}
\left[
\dfrac{\partial^2}{\partial r^2}
+
\dfrac{1}{\partial r}\dfrac{\partial}{\partial r}
-
\dfrac{m^2}{r^2}
+
K_m^2
\right]f(r)
=0, 
\label{r-fundeq2}
\end{eqnarray}
where 
\begin{eqnarray}
K_m^2 =n^2k^2-2k\dfrac{\Omega}{c}m.
\end{eqnarray}
Eq.~(\ref{r-fundeq2}) is the Bessel differential equation, 
and so the solution $f(r)$ should be the Bessel function 
of the m$^{\mbox {th} }$ order $J_m(K_m r)$.
The eigenvalue of the wave number $k$ is given by 
the zero of $J_m(K_m R)$ because of the Dirichlet boundary condition, 
where $R$ is the radius of the microdisk cavity. 

Accordingly, the shifted wave number due to rotation is 
\begin{eqnarray}
k = k_0 + \dfrac{\Omega}{n^2c}m 
+ 
O\left(
\left|
\frac{\Omega}{c}
\right|^2
\right),
\label{circular-wavenumber}
\end{eqnarray}
where $k_0$ is the zero of the Bessel function $J_m(nk R)$,  
that is, the eigenvalue of the wave number when the cavity 
is not rotating. 
Consequently, when the microdisk cavity is rotating, 
the wave function is the rotating wave $J_m(K_m r) e^{im\theta}$ 
and the degenerate wave number $k_0$ without rotation splits into 
two different wave numbers of the counter-propagating waves 
corresponding to the signs of the integer $m$. 
From Eq.~(\ref{circular-wavenumber}), one obtains 
the frequency difference $\Delta\omega$ between the counter-propagating waves,
\begin{eqnarray}
\Delta\omega = 2\dfrac{m}{n^2 }\Omega.
\end{eqnarray}

It is important that the CW and CCW wave solutions are 
degenerate eigenstates even when the microdisk cavity is not rotating, 
and hence the standing wave solution produced by the superposition of 
these degenerate rotating waves are also the eigenfunction. 
However, with a finite angular velocity $\Omega$, 
the CW and CCW wave solutions become 
non-degenerate states, which means that only the rotating   
waves are the eigenfunction of the rotating microdisk cavity. 
The frequency difference $\Delta \omega$ 
between the CW and CCW solutions 
is proportional to the angular velocity $\Omega$. 
This is the Sagnac effect for microdisk cavities. 

In general cases of 2D resonant microcavities of arbitrary shapes, 
Eq.~(2) cannot be directly solved in the same way for microdisks. 

First, we discuss the case that the spacing $\Delta k$ 
between the adjacent eigenvalues $k$ of the wave number is large enough to    
satisfy the following inequality,
\begin{equation}
\frac{\Omega}{c n^2 \Delta k} <
\left| 
\int \int_D d\rr \psi_0 
\left(
y\dfrac{\partial }{\partial x}-x\dfrac{\partial }{\partial y}
\right)
\psi_1 
\right|^{-1}, 
\label{ineq}
\end{equation}
where $\psi_0$ and $\psi_1$ are the wave functions 
of these eigenstates which correspond to the adjacent eigenvalues 
when the angular velocity $\Omega$ is zero.
We assume that, due to the rotation of the cavity, 
the eigenvalue is shifted as
$k=k_0 +\delta k$ and 
the wave function is changed as 
$
\psi = \psi_0 + 
\sum_{l \neq 0} c_{l}\psi_{l}, 
\label{expansion}
$
where 
$
\left(
\nabla^2 + n^2 {k_l}^2
\right)\psi_l
= 0. 
$
Here $\delta k$ and $c_l$ are assumed to be 
so small as $\Omega/c$.
Then, from Eq.~(\ref{fundeq}) we obtain 
\begin{eqnarray}
& 
\left[
2n^2 {k_0}^2\delta k
-2i k_0 
\left(\hh\cdot\nabla
\right)
\right]\psi_0
+
\sum_{l \neq 0}c_{l}n^2({k_0}^2-{k_l}^2)\psi_l
& \nonumber \\ 
& + 
O\left(
\left|
\dfrac{\Omega}{c}
\right|^2
\right)
= 
0.
& 
\label{eq32}
\end{eqnarray}
Using the following relation, 
\begin{equation}
\int\int_{D} d\rr
\psi_0 \left(\hh\cdot\nabla\right)\psi_0
=\frac{\Omega}{2c}\oint_{\partial D} {\psi_0}^2 \rr \cdot 
d {\mbox{\boldmath $s$}}
=0, 
\end{equation}
where $D$ and $\partial D$ denote respectively the domain 
and the edge of the cavity, 
we finally obtain $\delta k=0$ 
up to the first order of $|\Omega/c|$, and 
\begin{eqnarray}
c_l = 
\dfrac{2i {k_0} \int\int_{D} d\rr
\psi_l \left(\hh\cdot\nabla\right)\psi_0}
{ n^2({k_0}^2-{k_l}^2)}.
\label{cn}
\end{eqnarray}
Consequently, as long as the angular velocity is small,  
there is no Sagnac effect, which means 
that the wave functions are standing waves 
instead of counter-propagating waves 
and the frequency difference between two standing waves 
does not increase.

Next we discuss the case that the spacing $\Delta k$ between two 
wave numbers $k_0$ and $k_1$ is so small that 
it does not satisfy the inequality (\ref{ineq}). 
According to the perturbation theory for nearly-degenerate states 
in quantum mechanics, the wave function should be represented as 
the superposition of two nearly-degenerate eigenfuncions: 
$
\psi = c_0\psi_0 + c_1\psi_1.
$
Substituting this equation into Eq.~(\ref{fundeq}) yields 
\begin{equation}
M
\left(
\begin{array}{c}
c_0 \\
c_1
\end{array}
\right)=0,
\label{M}
\end{equation}
where 
$M$ is the following matrix: 
\begin{eqnarray}
\left(
\begin{array}{cc}
n^2(k^2-{k_0}^2) &
-2ik \int\int_{D} d\rr
\psi_0 \left(\hh\cdot\nabla\right)\psi_1\\
-2ik \int\int_{D} d\rr
\psi_1 \left(\hh\cdot\nabla\right)\psi_0 
& n^2(k^2-{k_1}^2)  
\end{array}
\right)
\nonumber 
\end{eqnarray}
In order to obtain non-trivial solutions for Eq.~(\ref{M}), 
the determinant of $M$ should vanish, which yields 
a quadratic equation for $k^2$. 
Consequently, we obtain the eigenvalues of 
the wave number up to the first order of $|\Omega/c|$, 
\begin{eqnarray}
k =\frac{k_0+k_1}{2}\pm \dfrac{1}{n^2}\left|\int\int_{D} d\rr
\psi_0 \left(\hh\cdot\nabla\right)\psi_1 \right|. \label{eq:k-sag}
\end{eqnarray}
Accordingly, the frequency difference $\Delta \omega$ 
between the two eigenfunctions newly produced by rotation of the cavity 
is proportional to the angular velocity\cite{Correspondence}, 
\begin{equation}
\Delta \omega = 2
\left|\int\int_{D} d\rr
\psi_0 \left(y\frac{\partial}{\partial x}
-x\frac{\partial}{\partial y}\right)\psi_1 \right|
\frac{\Omega}{n^2}. \label{eq:w-sag}
\end{equation}
Then, from Eq.~(\ref{M}) we also obtain 
the ratio of the coefficients $c_0$ and $c_1$ as follows:
\begin{equation}
c_1=\mp i c_0.
\label{cw}
\end{equation}

It is important that the standing wave function $\psi_0$ without rotation 
of the cavity can be represented 
by the superposition of the Bessel functions as 
$\psi_0 = \sum_{m=0}^{\infty} a_m J_m(n k_0 r) \cos{m\theta}$ 
, 
if the cavity is symmetric for $x$-axis and $\psi_0$ is an even function. 
The other wave function $\psi_1$ nearly degenerate to $\psi_0$ 
which should be an odd function for $x$-axis also can be written as 
$\psi_1 = \sum_{m=0}^{\infty} b_m J_m(n k_1 r) \sin{m\theta}$ 
, 
where $a_m\sim b_m$ and $k_0\sim k_1$.
When the cavity is rotating but the angular velocity is small enough 
to satisfy the inequality (\ref{ineq}), the frequency difference 
does not increase and the wave functions do not change drastically.
However, when the angular velocity is increased to be larger than 
this threshold which violates the inequality (\ref{ineq}),  
the wave functions change into the CW and CCW rotating waves 
because of Eq.~(\ref{cw}),
\begin{eqnarray}
\psi &=& \sum_{m=0}^{\infty} a_m J_m(n k r) 
\left(\cos{m\theta}\mp i\sin{m\theta}\right)
\nonumber \\ 
&=&\sum_{m=0}^{\infty} a_m J_m(n k r) e^{\mp im \theta}, 
\end{eqnarray}
where the wave number $k$ is given by Eq.~(\ref{eq:k-sag}).
Then, according to Eq.~(\ref{eq:w-sag}), 
one can observe the Sagnac effect as the frequency difference 
$\Delta \omega$
between counter-propagating waves proportional to the angular 
velocity.
The most important point is that there is a threshold of 
the angular velocity for the Sagnac effect in resonant microcavities.
It is possible to delete this threshold by symmetric shapes of 
the cavities as shown in the case of microdisks.

Next we numerically solve the Maxwell equation for a rotating microcavity,  
and show that the threshold phenomena of the transition from 
the standing wave solutions to the counter-propagating wave ones 
actually occur. 

First, let us explain the numerical method briefly. 
The solutions $\psi(r,\theta)$ for the Maxwell Equation (\ref{r-fundeq})
for the stationary solutions of the rotating cavity 
can be represented by the superposition of the Bessel functions 
in the cylindrical coordinates, 
\begin{eqnarray}
\psi(r,\theta)=\sum_{m=-\infty}^{\infty}a_m J_m(K_mr)e^{im\theta}. 
\label{eq:num}
\end{eqnarray}
Then the Dirichlet boundary condition is imposed as 
$\psi(R(\theta),\theta)=0$, 
where $R(\theta)$ denotes the edge of the cavity.
Accordingly,
%
%
%
we obtain 
\begin{eqnarray}
\sum_{m=-\infty}^{\infty} 
\left(
\int_{0}^{2\pi }J_m(K_mR(\theta))e^{i(m-n)\theta}d\theta
\right) a_m 
\label{inteq}
= 0.
\end{eqnarray}
For numerical computation, the infinite sum over $m$ can be approximated 
by the finite sum from $-M$ to $M$ where $M$ is a large integer. 
Then Eq.~(\ref{inteq}) can be rewritten as the matrix form, 
\begin{eqnarray}
&
\left(
\begin{array}{cccc}
A_{-M-M}  
& \cdots & A_{-MM} \\
\vdots     
&\ddots  & \vdots \\
A_{M-M}    
& \cdots & A_{MM} \\
\end{array}
\right)
\left(
\begin{array}{c}
a_{-M} \\
\cdots \\
a_{M}
\end{array}
\right) =0,
\end{eqnarray}
where 
$A_{nm}=\int_{0}^{2\pi }J_m(K_mR(\theta))e^{i(m-n)\theta}d\theta.$
Therefore, the eigenvalues of the wave numbers can be obtained 
numerically as the zeros of the determinant of the above matrix 
because of non-trivial solutions for the coefficients $a_m$. 

For numerical calculation, we choose a quadrupole cavity,
which is defined by 
the boundary $R(\theta)= R_0(1+\epsilon \cos 2\theta)$\cite{Nockel Stone}.
The parameters of the quadrupole are set as follows:
$R_0 = 6.2866\mu m, \epsilon = 0.12$,
and the refractive index $n = 1$.
When the quadrupole cavity is not rotating, 
solving the Helmholtz equation Eq.~(3) with $\Omega=0$ 
yields the nearly-degenerate standing wave eigenfuctions 
as shown in Fig.~1. We call the two modes shown in Fig.~1(a) and (b) 
modes A and B, respectively.

\begin{figure}
\begin{center}
  \begin{tabular}{ c c }
\raisebox{0.0cm}{\includegraphics[width=3.2cm]{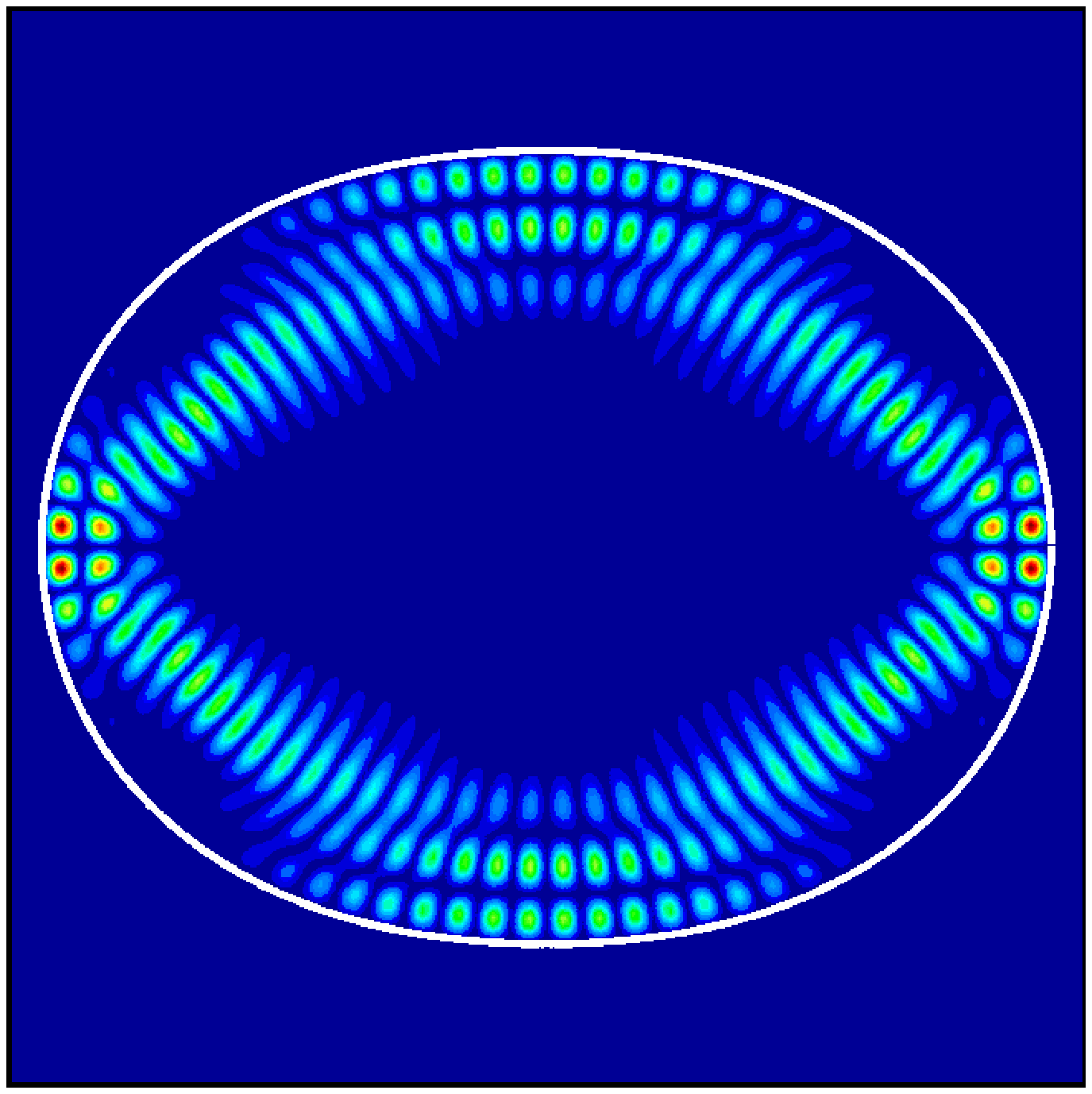}}
    &
    \includegraphics[width=3.2cm]{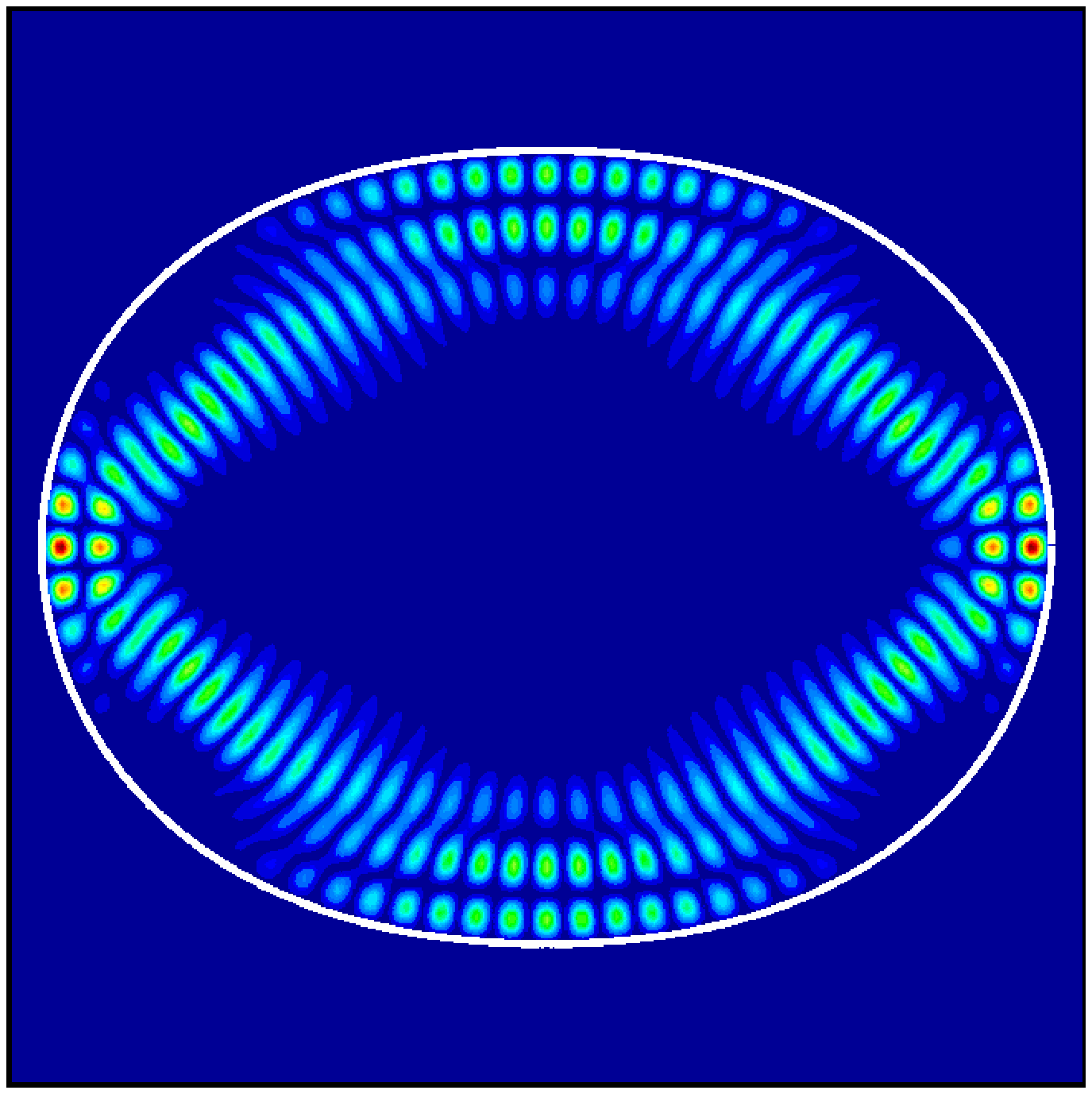}\\
    (a) & (b) \\
  \end{tabular}
\end{center}
\vspace{-4mm}
\caption{\label{fig1} 
The nearly-degenerate wave functions of the non-rotating quadrupole cavity 
corresponding to the eigen-wave numbers (a)$nk_AR_0=49.3380585$ and 
(b)$nk_BR_0=49.3380615$.
We call the modes of (a) and (b) modes A and B, respectively.
}
\end{figure}
%
\begin{figure}
\begin{center}
\raisebox{0.0cm}{\includegraphics[width=5.5cm]{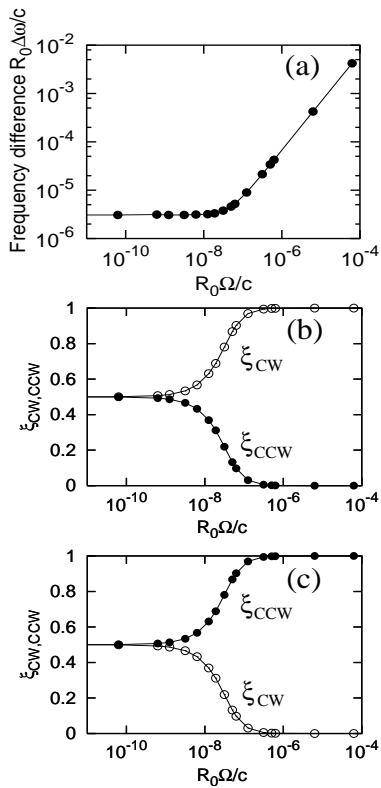}} \\
\end{center}
\vspace{-8mm}
\caption{\label{fig4}
(a)The (dimensionless) frequency difference 
$R_0\Delta\omega/c$ 
versus 
the (dimensionless) angular velocity.
The frequency difference does not change for 
$R_0\Omega/c < R_0\Omega_{th}/c (=5.0\times 10^{-8})$.
For $R_0\Omega/c >R_0\Omega_{th}/c$, 
the frequency difference becomes proportional to the angular velocity.
The CW and CCW waves components versus the angular velocity 
are shown for (b) mode A and (c) mode B.
}
\end{figure}

When the angular velocity $\Omega$ is smaller than 
a certain threshold $\Omega_{th}$  
(where $ R_0\Omega_{th}/c = 5.0\times 10^{-8}$), 
the frequency difference does not increase as shown in Fig.~2(a), 
and the eigenfunctions remain standing waves.
However, for $\Omega > \Omega_{th}$,
the frequency difference increases gradually,
and becomes proportional to $\Omega$, and 
modes A and B drastically 
change into the rotating wave functions 
as shown in Fig.~3(a) and (b), respectively.

The transition from the standing wave to the rotating wave 
can be clearly observed also by the CW and CCW wave components 
$\xi_{CW,CCW}$ defined as follows based on 
rotating wave decomposition of Eq.~(\ref{eq:num}):
\begin{equation}
\xi_{CW(CCW)} \equiv \sum_{m<0(m>0)} |a_m|^2/\xi, 
\end{equation}
where $\xi \equiv \sum_{m\neq 0} |a_m|^2$.
For $\Omega < \Omega_{th}$, $\xi_{CW,CCW}$ are around 0.5 
for modes A and B as shown in Fig.~2(b) and (c). 
When $\Omega$ exceeds $\Omega_{th}$, 
$\xi_{CCW(CW)}$ of mode A(B) suddenly vanishes, which means 
the wave function consists of only CW(CCW) waves. 
Therefore, one can see that modes A and B in Fig.~3 are 
the CW and CCW rotating waves, respectively. 

In summary, we have shown that 
the Sagnac effect can be observed in 2D resonant microcavities
when the angular velocity is larger than a certain threshold
where the nearly degenerate standing wave eigenfunctions of 
the non-rotating cavity change into the pair of 
the counter-propagating waves.

The threshold phenomenon seems to be akin to the lock-in phenomenon, which 
occurs owing to the mode-locking between the counter-propagating waves\cite{Crow,Aronowitz,Lamb2}.
However, in our theoretical approach, the effects of backscattering
and an active medium which both cause the lock-in phenomenon are not
taken into account.
The existence of the threshold can be shown even without these effects.

The Sagnac effect in microcavities will be actually observable 
by measuring the frequency difference in a rotating frame of reference,
as the measuring method of optical gyroscopes \cite{Crow,Aronowitz}.
A discussion on the actual experiment will be reported elsewhere.

\begin{figure}
\vspace{5mm}
\begin{center}
  \begin{tabular}{ c c }
 \raisebox{0.0cm}{\includegraphics[width=3.2cm]{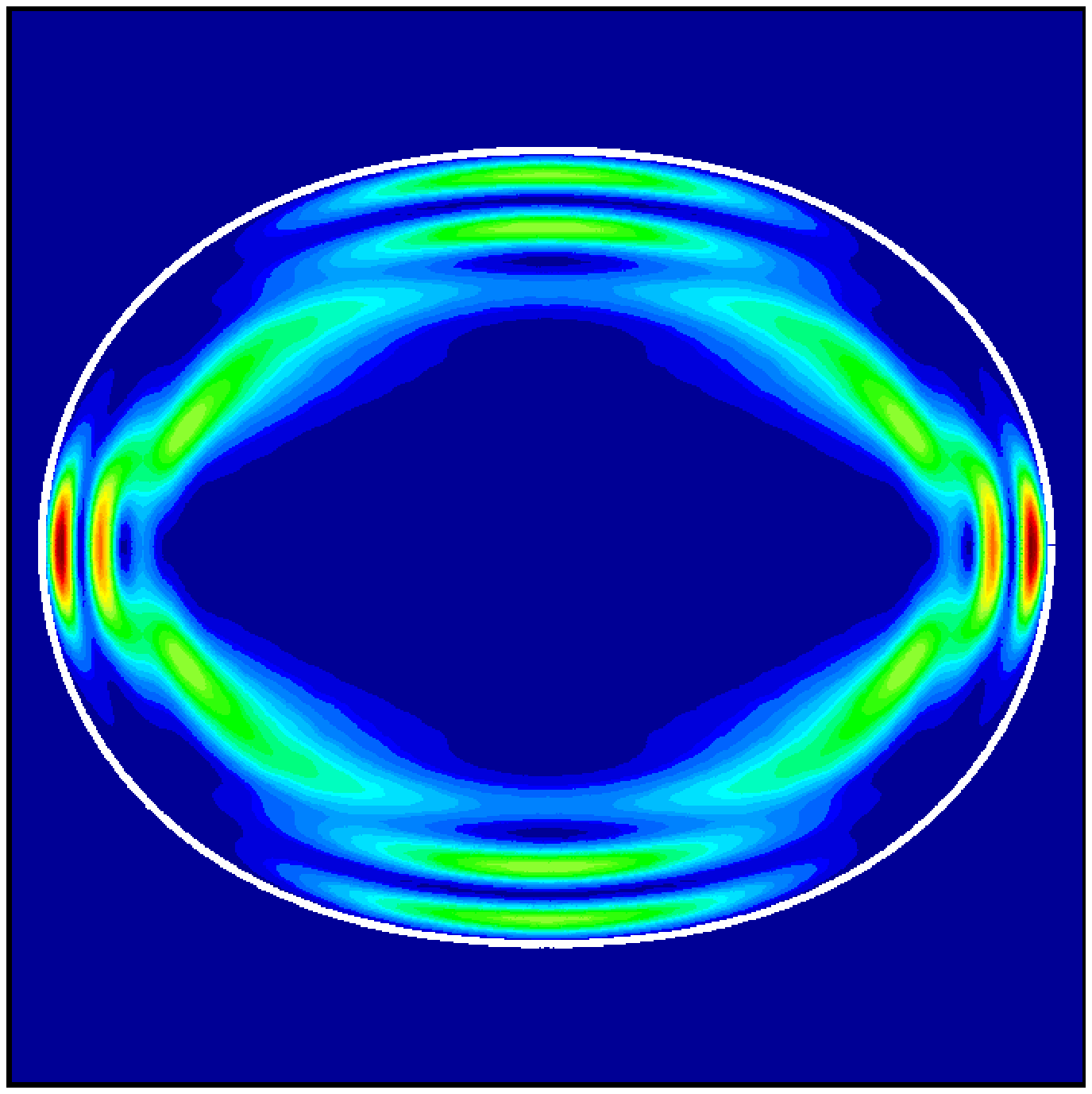}}
    &
    \includegraphics[width=3.2cm]{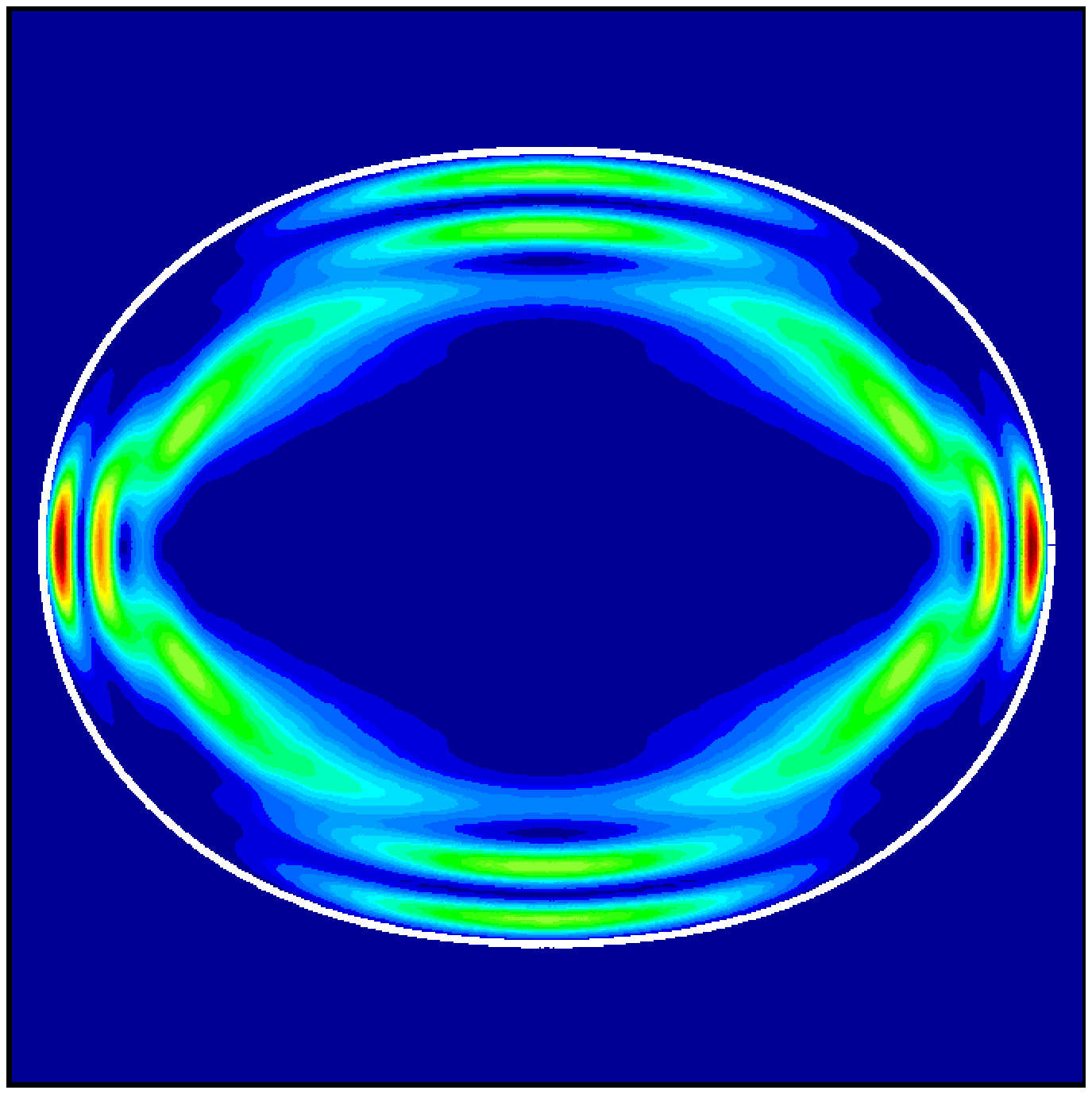}\\
    (a) & (b) \\
  \end{tabular}
\end{center}
\vspace{-5mm}
\caption{\label{fig3} 
The wave functions of the rotating quadrupole cavity 
with the (dimensionless) angular velocity  
$R_0\Omega/c= 6.28\times 10^{-7} (> R_0\Omega_{th}/c)$
respectively corresponding to (a)mode A and (b)mode B.
}
\end{figure}


\begin{acknowledgments}
The work was supported by 
the National Institute of information and Communication Technology
of Japan. 
\end{acknowledgments}


\begin{thebibliography}{99}
\bibitem{Correspondence} 
Eq.~(\ref{eq:w-sag}) can reproduce a conventional expression 
for the frequency difference: $(4A\omega)/(cnP)\Omega$ by 
assuming that the light propagates one-dimensionally and the 
resonant modes in a non-rotating microcavity 
can be described as one-dimensional WKB solutions along an optical
	path. 
(In the above expression, A is the area bounded by the optical path and 
$P$ is the perimeter of the optical path.)
\end{thebibliography}
\end{document}